\documentclass[12pt]{article}
\renewcommand{\title}[1]{\begin{center}\bf\Large #1\end{center}}
\renewcommand{\author}[1]{\begin{center}\large #1\end{center}}
\usepackage{latexsym,amsfonts}
\newcommand{\rr}{\mathbb{R}}

\begin{document}

\title{A Causal Algebra for Liouville Exponentials}

\vspace{7mm}

\author{
{\bf C. Ford}${}^a$  ~{\small and } {\bf G. Jorjadze}${}^b$ \\
\vspace{5mm}
{\small  ${}^a$School of Mathematics,}\\
{ \small Trinity College, Dublin 2, Ireland}\\
{ \small\tt ford@maths.tcd.ie}\\
\vspace{3mm}
{\small ${}^b$Razmadze Mathematical Institute,}\\
{\small M.Aleksidze 1, 0193, Tbilisi, Georgia}
\\
{ \small\tt jorj@rmi.acnet.ge }}

\vspace{7mm}

\begin{abstract}

\noindent {A causal Poisson bracket algebra for Liouville
exponentials on a cylinder is derived using an exchange algebra for
free fields describing the \it in \rm\, and \it out \rm\,
 asymptotics. The causal
 algebra
involves an even number of space-time points with a minimum of four.
 A quantum
realisation of the algebra is obtained which preserves causality and
the local form of non-equal time brackets.}
\end{abstract}

\vspace{10mm}

\baselineskip=18pt

Following Polyakov's work on the relativistic string \cite{P} the
quantum Liouville theory became the subject of intense study.
Indeed, the quantisation of Liouville theory is a deep problem in
its own right. Early approaches \cite{GN,CT,BCT,D'H J,OW} were based
on the canonical quantisation of structures present
in the classical
Liouville theory. In particular, exact forms were obtained for
operators corresponding to negative integer and half-integer powers
of the Liouville exponential. For arbitrary powers  a formal power
series can be obtained \cite{OW,KN}.

 In the 1990's
attention was focussed on the construction of correlation functions.
Dorn and Otto \cite{DO} and Zamolodchikov and Zamolodchikov
\cite{ZZ}  provided formulae for three-point functions and gave a
self-consistent framework for computing $n$-point functions.
However, an explicit construction of arbitrary Liouville
 exponential
operators is lacking (see however \cite{T, JW1}). In \cite{JW} it
was noted that the Liouville exponential obeys a  causal algebra at
the classical and quantum level. The Poisson bracket  of Liouville
exponentials at different space-time points can be expressed in
terms of the  exponential at four points (the two original points
plus two more where the light cones emanating from
 the two
points intersect). However, for exponentials defined on a cylinder
this four-point relation is only valid if the time separation is
sufficiently small. In this Letter a closed  causal algebra is
derived for all time separations. The number of points depends on
the temporal and spatial separations. The construction is based on
an exchange algebra for the free fields describing the \it in \rm\,
and \it out \rm\,
asymptotics of the Liouville field. A quantum
version of the exchange algebra is obtained through canonical
quantisation. This leads to a quantum realisation of the causal
algebra for the Liouville exponential to the power $-\frac{1}{2}$.

Classically, the Liouville field, $\varphi(x,\bar x)$, satisfies the
equation of motion
 \begin{equation}\label{motion}
\partial_x\partial_{\bar
x}\varphi(x,\bar x) +\mu^2e^{2\varphi(x,\bar x)}=0.
\end{equation}
Here $x=\tau+\sigma$ and $\bar x=\tau-\sigma$ are chiral coordinates
($ \tau$ is time and $\sigma$ is the spatial coordinate) and
 $\mu$ is a constant.
 The Liouville field
can be built out of canonical free fields. The simplest object is
the exponential in the equation of motion, $e^{2\varphi(x,\bar x)}$,
to the power $-\frac{1}{2}$
\begin{equation}\label{exponential}
V(x,\bar x)=e^{-\varphi(x,\bar x)}= E(x,\bar x)+F(x,\bar
x),\end{equation}
 where $E$
and $F$ are exponentials of massless free fields
\begin{equation}\label{EandF}
E(x,\bar x)=e^{-\varphi_{in}(x,\bar x)}, ~~~~~~~ F(x,\bar x)
=e^{-\varphi_{out}(x,\bar x)}.\end{equation}
 The free fields $\varphi_{in}$ and $\varphi_{out}$
 match the Liouville field $\varphi$
 in the limits $\tau\rightarrow
 -\infty$ and $\tau\rightarrow \infty$, respectively.

We now specialise to a cylindrical spacetime and
 take the
circumference to be $2\pi$ so that $\sigma\equiv \sigma+2\pi$. The
free field $\varphi_{in}$ has a standard Fourier expansion
\begin{equation}\label{FF}
\varphi_{in}(\tau,\sigma)=q+\frac{p\tau}{2\pi}
 +\frac{i}{\sqrt{4\pi}}\,\sum_{n\neq 0}
\left( \frac{a_n}{n}\, e^{-inx}+\frac{\bar a_n}{n}\ e^{-in\bar
x}\right),
\end{equation}
and the non-zero Poisson brackets read
\begin{equation}\label{C-C-R}
\{q,p\}=1,~~~~ \{a_m, a_n\} =\{\bar a_m,\bar a_n\}
=\,-im\,\delta_{m+n,\,0}.
\end{equation}
 The Fourier modes of $\varphi_{out}$ are related to
those of $\varphi_{in}$ through a canonical transformation. This is
the classical analogue of the unitary transformation which defines
the quantum $S$ matrix. The canonical transformation is most
conveniently expressed as a formula for $F(x,\bar x)$
\begin{equation}\label{Fdef}
F(x,\bar x)= \frac{\mu^2 e^{-\varphi_{in}(x,\bar x)}}{ 4\sinh^2
\frac{1}{2} p} \int^{2\pi}_0 dz \int^{2\pi}_0d\bar z ~
e^{\frac{1}{2} p[\epsilon(x-z) +\epsilon(\bar x-\bar z)]}
e^{2\varphi_{in}(z,\bar z)},
\end{equation}
where $\epsilon$ denotes the stairstep function defined by
\begin{equation}
\epsilon(x)=2n+1~~~~\mbox{ for } ~~~~ 2\pi n<x<2\pi(n+1).
\end{equation}

 The field
$\varphi(x,\bar x)=-\log V(x,\bar x)$ satisfies the field equation
(\ref{motion}) and defines a canonical map exchanging the Liouville
and \it in \rm\, fields \cite{D'H J,OW,J,FS}. The transformation is
not defined for $p=0$, and to obtain a one to one correspondence the
sign of $p$ must be fixed. Our interpretation of (\ref{FF}) as the
\it in \rm\, field is consistent with $p>0$; if $p>0$ $F(x,\bar
x)\rightarrow0$ as $\tau\rightarrow -\infty$ so that $V\sim
E=e^{-\varphi_{in}}$ in this limit.

A causal algebra for $V$ can be derived through the exchange algebra
satisfied by the $E$ and $F$ fields. The classical algebra reads
\begin{eqnarray}\label{PBA1}
\{E(x,\bar x),E(y,\bar y)\}&=&-\frac{1}{4}\,E(x,\bar x)E(y,\bar y)
\Bigl(\epsilon(x-y)+\epsilon(\bar x-\bar y)\Bigr) ,
\\
\nonumber \{E(x,\bar x),F(y,\bar y)\}&=&\frac{1}{4}E(x,\bar
x)F(y,\bar y) \Bigl(\epsilon(x-y)+\epsilon(\bar x-\bar y) \Bigr)
\\ \nonumber
 &&~+\frac{1}{2}
E(y,\bar x) F(x,\bar y)
 \frac{e^{-\frac{1}{2} p\epsilon(x-y)}}{
\sinh \frac{1}{2} p}
 +\frac{1}{2} E(x,\bar y)F(y,\bar x)
\frac{e^{-\frac{1}{2} p\epsilon(\bar x-\bar y)}}{\sinh \frac{1}{2}
 p},\\
 \nonumber \{F(x,\bar x),F(y,\bar y)\}&=&-\frac{1}{4}F(x,\bar
x)F(y,\bar y)\Bigl(\epsilon(x-y)+\epsilon(\bar x-\bar y) \Bigr).
\end{eqnarray}
  Using the Fourier expansion (\ref{FF}) the $\{E, E\}$ bracket is
easily derived. The derivation of the other two is outlined in the
Appendix.
The exchange
 algebra provides the non-equal time Poisson bracket of
the Liouville field exponential $V$
 in terms of $E$ and $F$.
The result
 simplifies if $x-y$ and $\bar x-\bar y$ are in the
`fundamental' domain $(-2\pi,2\pi)$. This can always be assumed if
the temporal separation is less than $\pi$.
 In this case
$\,\epsilon(x-y)=\mbox{sign}\,(x-y)\,$
 and $e^{-\frac{1}{2} p\epsilon(x-y)}=
\cosh \frac{1}{2} p-\sinh \frac{1}{2} p~\mbox{sign}(x-y)$ which
yields
\begin{eqnarray}\label{VV}
\{V(x,\bar x),V(y,\bar y)\}&=&-\frac{1}{4} \Bigl[E(x,\bar x)E(y,\bar
y)+ F(x,\bar
x)F(y,\bar y) +2E(x,\bar y)F(y,\bar x) \Bigr.\\
\nonumber&&~~~\Bigl. +2E(y,\bar x)F(x,\bar y)-E(x,\bar
x)F(y,\bar y)-E(y,\bar y)F(x,\bar x)\Bigr]\\
\nonumber &&~~~~\times~\Bigl(\mbox{sign}(x-y)+\mbox{sign}(\bar
x-\bar y) \Bigr).
\end{eqnarray}
Causality follows since $\mbox{sign}(x-y)+\mbox{sign}(\bar x-\bar
y)=0$ if $(x-y)(\bar x-\bar y)<0$. Using $E(x,\bar y)E(y,\bar
x)=E(x,\bar x)E(y,\bar y)$ and a similar formula for $F$, the right
hand side of (\ref{VV}) can be expressed in terms of $V$ at the four
points $(x,\bar x)$, $(y,\bar y)$, $(x,\bar y)$ and $(y,\bar x)$
  \cite{JW}
\begin{eqnarray}\label{VV1}
\{V(x,\bar x),V(y,\bar y)\}&=&\frac{1}{4}\left[ V(x,\bar x)V(y,\bar
y) -2V(x,\bar y)V(y,\bar x)\right]
\\ \nonumber
&&~~~~\times~
\Bigl(\mbox{sign}(x-y)+ \mbox{sign}(\bar x-\bar y)\Bigr).
\end{eqnarray}

Outside the fundamental domain the situation is more complicated.
Using the identity
\begin{equation}\label{thetasum}
\frac{e^{-\frac{1}{2}p\epsilon(x)}}{\sinh \frac{1}{2} p}=
\sum^\infty_{m=-\infty}e^{mp}\Bigl( 1-\mbox{sign}(x+2\pi m)\Bigr)
~~~~~~(p>0).
\end{equation}
and the shift property
\begin{equation}
E(y,\bar x-2\pi m)F(x+2\pi m,\bar y)=e^{mp}E(y,\bar x) F(x,\bar y),
\end{equation}
the $\{E,F\}$ bracket in (\ref{PBA1}) can be rewritten as
\begin{eqnarray}\label{PBA2}
\{E(x,\bar x),F(y,\bar y)\}&=&\frac{1}{4} E(x,\bar x)F(y,\bar y)
\Bigl(\epsilon(x-y)+\epsilon(\bar x-\bar y)\Bigr)
\\
\nonumber &&~+\frac{1}{2}\sum_{n=-\infty}^\infty E(y,\bar x-2\pi
n)F(x+2\pi n,\bar y) \Bigl( 1-\mbox{sign}(x-y+2\pi n) \Bigr)\\
\nonumber &&~+\frac{1}{2} \sum_{n=-\infty}^\infty E(x+2\pi n, \bar
y) F(y,\bar x-2\pi n)\Bigl(1-\mbox{sign}(\bar x-\bar y-2\pi n)
\Bigr).
\end{eqnarray}
This leads to the general non-equal time Poisson bracket
\begin{eqnarray}\label{VVgeneral}
\{V(x,\bar x),V(y,\bar y)\}&=&\frac{1}{4} \sum_{n=-\infty}^\infty
\Bigl[ V(x,\bar x)V(y,\bar y)-2V(y,\bar x-2\pi n)V(x+2\pi n,\bar y)
 \Bigr]
\\ \nonumber
&&~~~~~~~~~~~~\times~ \Bigl(\mbox{sign}(x-y+2\pi n)+\mbox{sign}(\bar
x-\bar y -2\pi n)\Bigr).
\end{eqnarray}
Note that only a finite number of terms in the sum contribute (in
the fundamental domain only the $n=0$ entry); for sufficiently large
$|n|$ the points $(x+2\pi n,\bar x-2\pi n)$ and $(y,\bar y)$ have
spacelike separation.

The quantum Liouville theory can be accessed  through a canonical
quantisation of the $in$-field $\varphi_{in}$ in the Hilbert space
$L^2(R_+)\otimes {\cal F}$, where $L^2(R_+)$ corresponds to the
momentum representation of zero modes and ${\cal F}$ is the standard
Fock space for the oscillator modes. The brackets (\ref{C-C-R}) are
replaced with the commutators
\begin{equation}\label{p,q}
[q,p]=i\hbar,~~~~~[a_m,a_n]=[\bar a_m,\bar
a_n]=\hbar\,m\,\delta_{m+n,\,0}.
\end{equation}
Our aim is to obtain quantum realisations of the exchange algebra
(\ref{PBA1}) and the causal algebra (\ref{VVgeneral}). To that end
we require operators corresponding to the exponentials $E$ and $F$.
For $E$ one can simply take a normal ordered version of the
classical formula (\ref{EandF})
\begin{equation}\label{quantumE}
E(x,\bar x)=:e^{-\varphi_{in}(x,\bar x)}:.
\end{equation}
The \it in \rm field representation of $F$ is deformed at the
quantum level
 \cite{OW}
\begin{equation}\label{quantumF}
F(x,\bar x)=\mu^2\int^{2\pi}_0dz\int^{2\pi}_0d\bar z~ :
\frac{e^{\frac{1}{2}p[\epsilon(x-z)+\epsilon(\bar z-\bar x)]}
}{4(\sinh^2
\frac{1}{2}p+\sin^2\frac{1}{4}\hbar)}e^{-\varphi_{in}(x,\bar x)}
e^{2\varphi_{in}(z,\bar z)}:f(x-z)f(\bar x- \bar z).
\end{equation}
 Here $f$ is a
short-distance factor
\begin{equation}
f(x)=\left(4\sin^2\frac{x}{2}\right)^{\hbar/(4\pi)}.
\end{equation}
 As usual normal
ordering is defined by placing creation ($n<0$ modes in (\ref{FF}))
and annihilation operators ($n>0$ modes) to the left and right,
respectively. Hermitian normal ordering of the zero modes is assumed
\cite{BCT,OW}, i.e. $:e^{2q}g(p):=e^q g(p) e^q$.

 A quantum version of the exchange algebra can be derived by
computing the operator products $E(x,\bar x)E(y,\bar y)$, $F(x,\bar
x)F(y,\bar y)$, $E(x,\bar x)F(y,\bar y)$ and relating these to
products of operators defined at different points.
 The simplest of
these is the $E\cdot E$ product which satisfies
\begin{equation}\label{quantumEE}
\frac{E(x,\bar x)E(y,\bar y)}{ e^{-\frac{1}{4}i\hbar \Theta(x-y,\bar
x-\bar y) }}- \frac{E(y,\bar y)E(x,\bar x)}{e^{\frac{1}{4}i\hbar
\Theta(x-y,\bar x-\bar y)}}=0,
\end{equation}
where
\begin{equation}
\Theta(x,\bar x)=\frac{\epsilon(x)+\epsilon(\bar x)}{2}.
\end{equation}
This is the quantum version of the $\{E,E\}$ bracket; expanding in
powers of $\hbar$ yields $[E(x,\bar x),E(y,\bar y)]= -\frac{1}{2}
 i\hbar E(x,\bar
x)E(y,\bar y)\Theta(x-y,\bar x-\bar y)+O(\hbar^2)$. The $F$ operator
satisfies the same commutation relation
\begin{equation}\label{quantumFF}
\frac{F(x,\bar x)F(y,\bar y)}{e^{-\frac{1}{4}i\hbar \Theta(x-y,\bar
x-\bar y)}}- \frac{F(y,\bar y)F(x,\bar x)}{ e^{\frac{1}{4} i\hbar
\Theta(x-y,\bar x-\bar y)}}=0.\end{equation} This follows from the
assumption that $E$ and $F$ are related by a unitary transformation,
i.e. the existence of the $S$ matrix. It can also be derived
directly from   (\ref{quantumF}). The quantum analogue of the
$\{E,F\}$ bracket is
\begin{eqnarray}\label{qexchangeEF}
\frac{E(x,\bar x)F(y,\bar y)}{ e^{\frac{1}{4}i\hbar \Theta(x-y, \bar
x-\bar y)}}&-&\frac{F(y,\bar y)E(x,\bar x)}{ e^{-\frac{1}{4}i\hbar
\Theta(x-y,\bar x-\bar y)}}
\\ \nonumber
&=& \frac{i\sin \frac{1}{2}\hbar}{ 2\sinh \frac{1}{2}p}\left[
e^{-\frac{1}{2}p\epsilon(x-y)} \left( \frac{E(y,\bar x)F(x,\bar y)}{
e^{-\frac{1}{4}i\hbar \Theta(x-y, \bar y-\bar x)}}+ \frac{F(x,\bar
y)E(y,\bar x)}{ e^{\frac{1}{4}i \hbar
\Theta(x-y,\bar y-\bar x)}}\right)\right.\\
\nonumber &&~~~~~~~~~~~\left.+ e^{-\frac{1}{2}p\epsilon (\bar x-\bar
y)} \left(\frac{E(x,\bar y)F(y,\bar x)}{ e^{\frac{1}{4}i\hbar
\Theta(x-y, \bar y-\bar x)}}+ \frac{F(y,\bar x)E(x,\bar y)}{
e^{-\frac{1}{4}i\hbar\Theta(x-y,\bar y-\bar x)}}\right)\right].
\end{eqnarray}
A derivation of this formula is given in the Appendix. Note that $p$
commutes with the product $E\cdot F$. The explicit $p$-dependence
can be removed using (\ref{thetasum}) and the quantum shift formula
\begin{equation}
\frac{E(y,\bar x-2\pi m)F(x+2\pi m,\bar y)}{
e^{-\frac{1}{4}i\hbar\Theta(x-y+2\pi m,\bar y-\bar x+2\pi m)}}=
e^{mp} \frac{E(y,\bar x)F(x,\bar y)}{
e^{-\frac{1}{4}i\hbar\Theta(x-y,\bar y-\bar x)}},
\end{equation}
giving a quantum form of (\ref{PBA2}).
This together with (\ref{quantumEE}) and (\ref{quantumFF})
 provides a causal algebra
for the operator $V(x,\bar x) =E(x,\bar x)+F(x,\bar x)$
\begin{eqnarray}\label{VVquantum}
&&\frac{V(x,\bar x)V(y,\bar y)}{ e^{\frac{1}{4}i\hbar\Theta(x-y,\bar
x-\bar y)}}- \frac{V(y,\bar y)V(x,\bar x)}{ e^{ -\frac{1}{4} i\hbar
\Theta(x-y,\bar x-\bar y)}}
\\
\nonumber &&~~~~=-\frac{i}{2} \sin \frac{\hbar}{2}
\sum_{n=-\infty}^\infty \left( \frac{V(y,\bar x-2\pi n)V(x+ 2\pi n,
\bar y)}{ e^{-\frac{1}{4} i\hbar \Theta(x-y+2\pi n,\bar y-\bar
x+2\pi n)}}+ \frac{ V(x+2\pi n, \bar y)V(y,\bar x-2\pi n)}{
e^{\frac{1}{4}i\hbar\Theta(x-y+2\pi n, \bar y-\bar x +2\pi
n)}}\right)
\\
\nonumber &&~~~~~~~~~~~~~~~~~~~~~~~~~~~~~\times \Bigl(
\mbox{sign}(x-y+2\pi n) +\mbox{sign} (\bar x-\bar y-2\pi n)\Bigr).
\end{eqnarray}
If $x-y$ and $\bar x-\bar y$ are in the interval $(-2\pi,2\pi)$ only
the $n=0$ summand contributes and $\Theta(x-y)$ reduces to
$\frac{1}{2}\left(\mbox{sign}(x-y)+ \mbox{sign}(\bar x-\bar
y)\right)$. In this case it is easy to recover
\begin{eqnarray}\label{VVcom}
[V(x,\bar x),V(y,\bar y)]=-i\,\sin\frac{\hbar}{4}\,
\Bigl(\mbox{sign}(x-y)+\mbox{sign}(\bar x -\bar y)\Bigr) \times
~~~~~~~~~~~~~~~~~~~~~~~~~~~~\\ \nonumber \left(V(x,\bar y)
 V(y,\bar
x)+V(y,\bar x) V(x,\bar y) - \frac{V(x,\bar x) V(y,\bar y)+V(y,\bar
y) V(x,\bar x)}{2\cos \frac{1}{4}\hbar}\right).
\end{eqnarray}
which was obtained  in \cite{JW} via Moyal quantisation.

In \cite{DO,ZZ} matrix elements of the form $\langle p| V(x,\bar
x)|p'\rangle$ were interpreted as three-point functions. The causal
algebra (\ref{VVquantum}) can similarly be understood as an identity
relating four-point functions. The approach followed here should
also be applicable to other integrable conformal field theories. The
 $SL(2,\rr)/U(1)$ black hole model has a similar free field
parameterisation \cite{FJW,K} and  quantum exchange and causal
algebras are expected here as well.

\vspace{7mm}

\noindent {\bf Acknowledgements~} We are grateful to Gerhard Weigt
and Harald Dorn for helpful
discussions. G.J. thanks the Humboldt
University, AEI in Potsdam and ICTP Triese for hospitality where
much of his work was done. His research was supported by grants from
the DFG, GRDF, INTAS, RFBR and GAS.

\setcounter{equation}{0}
\def\theequation{A.\arabic{equation}}

\newpage

\noindent{\bf \Large{Appendix }}

\vspace{3mm}

 Here we outline the derivation of the classical and quantum
exchange algebras. The canonical brackets (\ref{C-C-R}) are
equivalent to
\begin{equation}\label{phibracket}
\{\varphi_{in}(x,\bar x),\varphi_{in}(y,\bar y)\}= -\frac{1}{2}
\Theta(x-y,\bar x-\bar y).
\end{equation}
The $\{E,E\}$ bracket in (\ref{PBA1}) is an immediate consequence of
this formula. To derive the $\{E,F\}$ bracket it helps to write
 (\ref{Fdef})
as follows
\begin{equation}\label{Frep}
F(x,\bar x)=\frac{\mu^2}{ 4\sinh^2\frac{1}{2}p} E(x,\bar x)S(x,\bar
x),
\end{equation}
 and
\begin{equation}\label{sdef}
S(x,\bar x)= \int^{2\pi}_0dz\int^{2\pi}_0 d\bar z~ e^{
\frac{1}{2}p[\epsilon(x-z)+\epsilon(\bar z-\bar x)]}
e^{2\varphi_{in}(z,\bar z)}.
\end{equation}
Accordingly, we require the bracket of $E$ with $p$ and $S$, the
former is
\begin{equation}\label{Epbracket}
\{E(x,\bar x),p\}=-E(x,\bar x).
\end{equation}
For the $\{E,S\}$ bracket the following rewrite of (\ref{sdef}) is
useful
\begin{equation}\label{Srep}
S(x,\bar x)= \int^{2\pi}_0dz\int^{2\pi}_0~
 d\bar z ~e^{2
\varphi_{in}(x+z,\bar x+\bar z)-p},
\end{equation}
 which together with
(\ref{phibracket}) yields
\begin{eqnarray}\label{ESPB}
\{E(x,\bar x),S(y,\bar y)\}&=&\frac{1}{2}E(x,\bar x) \int^{2\pi}_0
dz \int^{2\pi}_0d\bar z~ e^{2\varphi_{in}(y+z,\bar y +\bar z)-p}\\
\nonumber &&~~~~~~~~~~~~~~~~\times
 \left[
\epsilon(x-y-z)+\epsilon(\bar x-\bar y-\bar z)+2\right].
\end{eqnarray}
Using the identity
\begin{equation}\label{eps-sh}
\epsilon(x-y-z)=\epsilon(x-y)-\epsilon(z)-\frac{\cosh \frac{1}{2}
p}{\sinh\frac{1}{2} p}+\frac{e^{\frac{1}{2}
p[\epsilon(x-y-z)-\epsilon(x-y)+\epsilon(z)]}}{\sinh \frac{1}{2} p},
\end{equation}
(\ref{Srep}) reduces to
\begin{eqnarray}\label{ESbracket}
\{E(x,\bar x),S(y,\bar y)\}&=&\, E(x,\bar x)S(y,\bar y)
\left(\Theta(x-y,\bar x-\bar y) -\frac{\cosh
\frac{1}{2}p}{\sinh\frac{1}{2}p} \right)\\ \nonumber&&~
  -\,\theta_{-p}(x-y)\,E(x,\bar x)S(x,\bar y)~
\,-\theta_{-p}(\bar x-\bar y)\,E(x,\bar x) S(y,\bar x),~~~~\,
\end{eqnarray}
where
\begin{equation}
\label{thetadef} \theta_p(x)=\frac{ e^{\frac{1}{2}p \epsilon(x)}}{
2\sinh \frac{1}{2}p}.
\end{equation}
This together with (\ref{Epbracket}) and the $\{E,E\}$ bracket leads
to the formula for $\{E,F\}$. To compute the $\{F,F\}$ bracket one
also requires the $\{S,S\}$ bracket.

In the quantum case
 (\ref{quantumEE}) follows from the commutator
\begin{equation}\label{varphicomm}
[\varphi_{in}(x,\bar x),\varphi_{in}(y,\bar y)] =-\frac{i\hbar}{2}
\Theta(x-y,\bar x-\bar y).
\end{equation}
Equation (\ref{qexchangeEF}) is less straightforward - this requires
the product of the $E$ and $F$ operators. The quantum $F$ defined in
equation (\ref{quantumF}) can be recast in the form
\begin{equation}
F(x,\bar x)=\mu^2 c(p)E(x,\bar x)S(x,\bar x),
\end{equation}
 where $E$ is as
in (\ref{quantumE}),
\begin{equation}
\label{cdef} c(p)=\frac{1}{4\sinh\frac{1}{2}p~ \sinh\frac{1}{2}
(p+i\hbar)},
\end{equation}
and
\begin{equation}\label{quantumSdef}
S(x,\bar x)= \int^{2\pi}_0 dz\int^{2\pi}_0 d\bar z :e^{2\varphi_{in}
(x+z,\bar x+\bar z)}:e^{-(p-i\hbar)}.
\end{equation}
  Note that the shift in $p$ renders $S$ hermitian.
 Although the product $E(x,\bar x)S(x,\bar x)$ involves
coincident points it is well defined; the product actually generates
the `short-distance' factors in the manifestly normal ordered form
(\ref{quantumF}). The normal ordering in (\ref{quantumSdef}) can be
avoided at the expense of performing a multiplicative
renormalisation of the $\mu$ parameter.

As a first step in computing the $E\cdot F$ product we consider the
product of $E(x,\bar x)$ and $S(y,\bar y)$; using (\ref{varphicomm})
the product can be written
\begin{eqnarray}\label{E-S}\nonumber
E(x,\bar x) S(y,\bar y)&=& \int_0^{2\pi}dz\int_0^{2\pi}d\bar z\,
:e^{2\varphi_{in}(y+z,\bar y+\bar z)}:
 e^{-(p-2i\hbar)} E(x,\bar x)
e^{\frac{1}{2} i\hbar[\epsilon(x-y-z)+\epsilon(\bar x-\bar y-\bar
z)]},\\
&&
\end{eqnarray}
which for $x=y$ and $\bar x=\bar y$ yields $E(x,\bar x) S(x,\bar x)=
S(x,\bar x) E(x,\bar x)$. This can be expressed as an exchange
relation using the identity (a generalisation of (\ref{eps-sh}))
\begin{equation}\label{eps-sh2}
\sinh \frac{p}{2} \,\,e^{\frac{1}{2}i\hbar[\epsilon(x-y-z)
-\epsilon(x-y)+\epsilon(z)]}=\sinh\frac{p-i\hbar}{2}+
i\sin\frac{\hbar}{2}\,\, e^{\frac{1}{2}
p[\epsilon(x-y-z)-\epsilon(x-y)+\epsilon(z)]}.
\end{equation}
The result reads
\begin{eqnarray}\label{E-S1}\nonumber
E(x,\bar x) S(y,\bar y)&=&
\frac{\sinh^2\frac{1}{2}p}{\sinh^2\frac{1}{2}(p+i\hbar)}\Bigl[
-\,2i\sin\frac{\hbar}{2}\,\,\theta_{-p}(x-y)\,e^{\frac{1}{2}i
\hbar\,\epsilon(\bar
x-\bar y)}\,S(x,\bar y) E(x,\bar x)\Bigr.\\
\nonumber
&&~~~~~~~~~~~~~~~~~~~~~-2i\sin\frac{\hbar}{2}\,\,\theta_{-p}(\bar
x-\bar y)\,e^{\frac{1}{2}i \hbar\,\epsilon(x-y)}\,S(y,\bar x)
E(x,\bar x)
\\ \nonumber
          &&~~~~~~~~~~~~~~~~~~~~~-4
          \sin^2\frac{\hbar}{2} \,\,\theta_{-p}(x-y)\,
\theta_{-p}(\bar x-\bar y)\,S(x,\bar x) E(x,\bar x)~~~\\
          &&~~~~~~~~~~~~~~~~~~~~~+\Bigl.
 e^{i\hbar\Theta(x-y,\bar x-\bar y)}S(y,\bar y)E(x,\bar x)\Bigr].
\end{eqnarray}
This leads to the operator product
\begin{eqnarray}\label{quantumEF}
E(x,\bar x)F(y,\bar y)&=& v(p)\Bigl[-2i\sin \frac{\hbar}{2}~
\theta_{-p}(x-y)e^{\frac{1}{2}i\hbar\epsilon(\bar x-\bar y)}
F(x,\bar y)E(x,\bar x)\Bigr.\\ \nonumber &&~~~~~~~-2i\sin
\frac{\hbar}{2}~\theta_{-p}(\bar x-\bar y)
e^{\frac{1}{2}i\hbar\epsilon(x-y)}F(y,\bar x)E(x,\bar x)\\
\nonumber &&~~~~~~~-4\sin^2\frac{\hbar}{2}~ \theta_{-p}(x-y)~
\theta_{-p}(\bar x-\bar y)F(x,\bar x)E(y,\bar y)
\\ \nonumber
          &&~~~~~~~~\Bigl.+e^{\frac{1}{2}i\hbar\Theta(x-y,\bar x-\bar y)}
          F(y,\bar y)E(x,\bar x)\Bigr],
\end{eqnarray}
where
\begin{equation}\label{vdef}
v(p)=\frac{c(p-i\hbar)}{c(p)} \frac{\sinh^2\frac{1}{2}p}
{\sinh^2\frac{1}{2}(p+i\hbar)}= \frac{\sinh^2\frac{1}{2}p}{
\sinh^2\frac{1}{2}p+\sin^2\frac{1}{2}\hbar}.
\end{equation}
 Equation (\ref{quantumEF}) and its hermitian conjugate can be
 manipulated
into (\ref{qexchangeEF}).

Finally,  (\ref{cdef}) can be used to write the $F$ operator in a
more symmetric fashion
\begin{equation}
F(x,\bar x)=\frac{\mu^2}{2\sinh\frac{1}{2}p} E(x,\bar x) S(x,\bar x)
\frac{1}{2\sinh\frac{1}{2} p}.\end{equation} In this form the
quantum deformations are
 hidden in the  operator products.

\end{document}